# A trustless society?
# A political look at the blockchain vision

Rainer Rehak

A lot of business and research effort currently deals with the so called decentralised ledger technology blockchain. Putting it to use carries the tempting promise to make the intermediaries of social interactions superfluous and furthermore keep secure track of all interactions. Currently intermediaries such as banks and notaries are necessary and must be trusted, which creates great dependencies, as the financial crisis of 2008 painfully demonstrated. Especially banks and notaries are said to become dispensable as a result of using the blockchain. But in real-world applications of the blockchain, the power of central actors does not dissolve, it only shifts to new, democratically illegitimate, uncontrolled or even uncontrollable power centers. As interesting as the blockchain technically is, it doesn't efficiently solve any real-world problem and is no substitute for traditional political processes or democratic regulation of power. Research efforts investigating the blockchain should be halted.

Numerous new blockchain research centers have recently been formed in universities all across Europe, the German federal government is working on a blockchain strategy and big blockchain conferences are being held worldwide. Having arrived at such level of attention, I want to take a closer look at the political implications of applying the blockchain technology to societally relevant services: Can this technology hold its general promises when practically applied or is it a hype actually being suitable for marginal use cases only? To guide this enquiry, I focus on understanding the political ramifications of a blockchain-based 'trustless' society in contrast to the current one.

Modern societies are based on trust: trust in other people, in certain procedures – such as democratic elections –, and in intermediary institutions such as banks. Without this generalized trust, complex societies based on the division of labour could not exist. Occasionally, however, this trust is fundamentally betrayed, and so it makes sense to look for new ways to minimize the need for trust in the societal coexistence. For that, I focus on the digital technology called blockchain, as applied in Bitcoin or Ethereum. The blockchain supposedly – if applied correctly – makes central intermediaries, so-called trusted third parties such as banks or notaries, superfluous. The blockchain should therefore solve the problem of creating a common consensus in a neutral technical-cryptographic way and no longer in an organisational way.

From a technical point of view, the blockchain is a mechanism for solving a problem which computer scientists call network consensus. The aim is to create and maintain a reliable common understanding concerning the current state of shared objects, even protected from manipulation. Classic examples of this problem are time synchronization, or the assignment of domain names to IP addresses. Every member of the network needs to have the same understanding – a common view – for the system to work. But the common logging of activities or monitoring value transactions are also instances of this problem. These do not only concern a current state, but also some kind of history containing past changes. So how can a network of distributed systems agree on what is currently "the case in the world"? The simplest solution to this problem is to have central authorities manage the task on behalf of the other systems involved and thus create coherence in a low-overhead and scalable manner. Ultimately however, in this setting, the power over all systems' common understanding is delegated to one or a few privileged central points which all parties must trust and which at best have no self-interest in manipulation. The blockchain attempts to prevent the concentration of power and vulnerability of those centralized approaches by technically forcing decentralisation, public reproducibility and immutability of data records. The blockchain therefore offers the functionality of a directory or ledger, but without (trusted) intermediaries, which is why the term "trustless" is often used.

How did this technology spread worldwide and can it keep the bold promises of its advocates in actual use cases? The blockchain-relevant topic of trust in intermediaries became crucial in 2008 due to the financial crisis and the resulting global recession. The key institutions – in this case, banks – had massively enriched themselves, manipulated relevant key figures, and thus produced a worldwide financial and trust fiasco, whose effects can still be seen globally today. However, banking regulations were neither significantly restricted, nor was a split-up of the big banks politically discussed. The banks were rescued with taxpayers' money and could essentially continue as before, while economies groaned worldwide and millions of people lost their savings, jobs, and homes. This (non-)reaction of state authorities to one of the most relevant events in recent economic history gave a huge boost to a subgroup within the critical tech community: the crypto-libertarians. They felt confirmed in their belief that any concentration of power does more harm than good – the only thing that counts is the free individual. From their perspective, one can and must use technical tools, in particular cryptographic ones, to defend against overbearing institutions. Even though this movement has existed for decades, its radical-individualist world-view began gaining traction outside its own ranks. In this situation, a person or group using the pseudonym Satoshi Nakamoto published a concept paper for an alternative global currency including its own means of payment – including usable software with the necessary cryptographic mechanisms. The crypto currency Bitcoin was born: a public distributed ledger without intermediaries securely documenting money transfers; in other words, a monetary system without banks.

Furthermore, this technology can not only be used for processing financial transactions, but also for recording any transaction of value or data and, due to the immutability of the transaction history, for reliably storing any information including executable programs. This abstraction and generalization of Bitcoin led to the blockchain technology, which can be utilised to process digital transactions, similar to a notary's office. In order to analyse the almost magical claim that a purely technically mediated and neutrally documented consensus can be achieved, I will briefly describe the mechanism of the blockchain. The blockchain consists of a chronologically ordered chain of data units, the so-called blocks, where each block contains a defined number of transactions and a cryptographically secure reference to the entire previous block. The whole blockchain thus contains the valid current state and the complete history of transactions in the network. A block cannot be changed afterwards, because otherwise the secure reference would be cut and following blocks would become invalid.

"The brilliance of the blockchain lies in the fact that all computers in the network concurrently try to form a new block using transactions not already stored in the blockchain. In effect, all computers independently try to solve a cryptographically complex task – a crypto puzzle." As soon as the first computer finds a solution it gets a (financial) reward. Immediately every other computer will stop working on the current block, and start creating the next block referencing the one just created by the winner – the chain just got extended. With Bitcoin, block creation happens approximately every ten minutes. The intention of the immense use of resources in the competition of parallel puzzle solving – the so-called mining – is that it is always another computer (equals: never the same) in the network creating a new block. So, if someone wanted to manipulate certain blocks to create an alternative history, that actor would have to be able to out-compute the rest of the network to stay ahead on every new block. This competitive model intends to prevent centralisation, hence allows for the system to work in a distributed manner. However, since the puzzle is a computationally intensive task, the probability of a solution increases with the computing power of the computer used.

Considering and contextualizing the technical properties, it is notable that a blockchain is technically decentralized, but – as commonly found in individualistic concepts – it assumes (actually requires) equally powerful actors. However, since generating a block in the Bitcoin network is financially rewarded, mining has been professionalised for several years now by merging computers and hardware specialisation. Commonly used laptops by private individuals now compete against storage building sized computing clusters equipped with highly customised graphics card chips – and practically always lose. This is because the computing power alone is decisive for winning the crypto puzzle and therefore successful block generation; it becomes obvious that the system has no internal mechanism to maintain its initial and intended decentralization. Depending on estimates, 50 to 60 percent of the so-called hash rate, i.e. the computing power of the entire Bitcoin network, is currently in the hands of the Chinese miner company Bitmain. This degree of centralization is comparable to or even greater than in the conventional banking system. In effect, in the largest active blockchain project decentralization is an illusion. Although the possibilities for manipulation are limited (for example, pending transactions can only be delayed or completely suppressed), the limitation explicitly does not originate from the system's (de)central character. It results from the publicly visible transaction records and the underlying asymmetric key cryptography for signing records. This kind of manipulation protection could also be easily achieved without using a blockchain. In addition, the blockchain understands decentralization only technically, not in terms of administrative power. So, if 90 percent of the individual computers in a network are under the control of a single person or organization, the network may still be distributed in a technical sense but concentrated from a power analysis perspective. This puts the expectations towards any blockchain-centric solution harshly into perspective.

The immutability analysis of the blockchain is also quite revealing from a political point of view. The concept of cryptographically secure logging is, in computer science terms, already ancient. It has been practically applied since at least the 1980s, for example in the form of *hash chains,* a way of storing data where every new record secures all previous ones. What is actually new with the blockchain is its distributed character, but this also has its own side effects. In 2016, for example, a programming error led to two parallel yet valid histories of the blockchain-based crypto currency Ethereum. To correct this fundamental defect, the two chains had to be laboriously reunited. Also, in 2016 a fully automated commercial organization called DAO (Decentralized Autonomous Organisation, a venture capital fund) was developed, but its code had bugs and was hacked. As a result, DAO was robbed of a third of its value in Etherium coins. The community could not agree on how to deal with this bug. Some wanted to fix the 'accident' and others wanted to stick to the immutability of history. Eventually the Ethereum blockchain was split up into one where the robbery had happened and one where it had not. Even Bitcoin itself had a blockchain split in 2017, as there was a dispute over technical parameters while extending the Bitcoin code basis. As a result, the parallel currency Bitcoin Cash was born.

In the light of those examples it turns out that the transaction log of a blockchain is technically unchangeable, but the usage contexts and social conditions of its real-world application have a great influence on the extent to which the technically implemented immutability is actually effective. Since the blockchain is software, changes of the kind previously described are caused by code changes. But not everyone can make changes to the code, so who decides in which direction the network is developed and how is it negotiated, if at all? If we state "code is law", then who is the "legislative"? Why are certain decisions implemented while others are not, and who does the proper implementing? Obviously, non-technical procedures – perhaps even quasi-democratic ones – for social negotiation, regulation, and conflict resolution play an essential role. But this was exactly what the blockchain originally wanted to make obsolete.

Another fundamental problem goes hand in hand with applying a blockchain to the physical world: Let us assume that a blockchain in fact ideally implemented the characteristics of immutability, publicity, distribution and thus, trustlessness: How could the correctness of the data stored in the blockchain be verified and ensured? Dealing with digital financial transactions is comparatively simple, since each person can only spend the money that is available in his/her own account. But as soon as it comes to claims about the physical world, such as whether cash or goods have been exchanged, or whether a piece property has been damaged, so that an insurance company would have to compensate, the blockchain only provides an immutable documentation of the allegations. The problem of correctness and reliability remains unsolved.

So far the eternal desire to solve social and societal problems through neutral technology remains unattainable, even with the blockchain. It does not make powerful intermediaries disappear, but only recreates them outside the scope of technology as can be observed with the advent of crypto currency exchanges centrally keeping many people's Bitcoins. Those new centres can be regulated and frequently betray their user, just like it was before. On the other hand, if the attempts to dissolve intermediaries were successful, they would plainly follow the neoliberal mantra of individualizing societal risks: Sole responsibility lies again on the shoulders of the individual person. Distressed are those who lose all their savings through a hack because their home computer and thus their Bitcoin wallet was not sufficiently secure. Distressed are those losing their retirement pay because the pension managing smart contract was poorly programmed. Maybe a small IT elite could profit from more freedom through blockchains, but the rest of society would most likely become more vulnerable; just like in medieval times of hiding one's money in bedsheets.

As a society, we have to make decisions: Is the illusion of getting along without trustworthy third parties really worth the massive resource expenditure of permanent parallel calculation? Certainly not. But can the foundation of society be reinforced by forging anti-institutionalism into technology? The answer to this question can be found in the Bitcoin blockchain: The absence of institutions counteracting power asymmetries ultimately leads to the anarcho-libertarian right of the (computationally) strongest. Those harsh results imply that public research on blockchain should be reduced to the few actual highly specialized technical use cases, while research funding for societal applications could and should be used much better elsewhere.

Ultimately, societal subsystems are always based on trust – the only relevant question is how trust can be negotiated and legitimized. Seen in this light, the Bitcoin project seems to be a well-hidden but very emphatic call for the overdue democratization of the banking system.

**Address and professional function of author:**
Rainer Rehak
Weizenbaum-Institute for the Networked Society
Hardenbergstraße 32
10623 Berlin
E-Mail: rainer.rehak@wzb.eu

Rainer Rehak is doctoral candidate in the research group "Quantification and Social Regulation" of the Weizenbaum-Institute for the Networked Society. His dissertation project deals with concepts for implementing and safeguarding democratic principles in IT systems.